\documentclass[english,aps,preprint]{revtex4}
\usepackage[T1]{fontenc}
\usepackage[latin9]{inputenc}
\usepackage{array}
\usepackage{float}
\usepackage{textcomp}
\usepackage{multirow}
\usepackage{amsmath}
\usepackage{graphicx}

\makeatletter

\providecommand{\tabularnewline}{\\}

\@ifundefined{textcolor}{}
{%
 \definecolor{BLACK}{gray}{0}
 \definecolor{WHITE}{gray}{1}
 \definecolor{RED}{rgb}{1,0,0}
 \definecolor{GREEN}{rgb}{0,1,0}
 \definecolor{BLUE}{rgb}{0,0,1}
 \definecolor{CYAN}{cmyk}{1,0,0,0}
 \definecolor{MAGENTA}{cmyk}{0,1,0,0}
 \definecolor{YELLOW}{cmyk}{0,0,1,0}
}

\makeatother

\usepackage{babel}
\begin{document}

\title{Resonant production of leptogluons at the FCC based lepton-hadron
colliders}

\author{Y. C. Acar}

\email{ycacar@etu.edu.tr}

\affiliation{TOBB University of Economics and Technology, Ankara, Turkey }

\author{U. Kaya}

\email{ukaya@etu.edu.tr}

\affiliation{TOBB University of Economics and Technology, Ankara, Turkey }

\affiliation{Department of Physics, Faculty of Sciences, Ankara University, Ankara,
Turkey}

\author{B. B. Oner}

\email{b.oner@etu.edu.tr}

\affiliation{TOBB University of Economics and Technology, Ankara, Turkey }

\author{S. Sultansoy}

\email{ssultansoy@etu.edu.tr}

\affiliation{TOBB University of Economics and Technology, Ankara, Turkey }

\affiliation{ANAS Institute of Physics, Baku, Azerbaijan}
\begin{abstract}
Resonant production of leptogluons at the FCC based ep and $\mu$p
colliders have been analyzed. It is shown that e-FCC and $\mu$-FCC
will cover much wider region of $e_{8}$ and $\mu_{8}$ masses than
the LHC. While leptogluons with appropriate masses (if exist) will
be discovered earlier by the FCC pp collider, lepton-proton colliders
will give opportunity to handle very important additional information.
For example, compositeness scale can be probed up to multi-hundred
TeV region.
\end{abstract}
\maketitle

\section{introduction}

Color octet leptons are predicted by preonic models (see \cite{key-1}
and references therein) with colored preons (see, for example, fermion-scalar
models \cite{key-2,key-3}). From phenomenological viewpoint their
status is similar to that of excited leptons and leptoquarks. Concerning
experimental searches situation is quite different: excited leptons
and leptoquarks occupy an important place in the research program
of almost all collider experiments, however, this is not the case
for leptogluons (see Chapter titled \textquotedblleft{}Quark and Lepton
Compositeness, Search for\textquotedblright{} in \cite{key-4} and
references therein). 

As for the phenomenological studies at TeV colliders: pair production
of leptogluons at the LHC have been considered in \cite{key-3,key-5,key-6,key-7}.
Resonant production of color octet electrons at the LHC based ep colliders
is analyzed in \cite{key-8,key-9,key-10}. In \cite{key-11} indirect
manifestations of color octet electrons at ILC and CLIC have been
considered. Resonant production of color octet muons at muon collider
based \textmu{}p colliders was considered in \cite{key-12}. It is
interesting that color octet neutrinos may be the source of the IceCube
PeV events \cite{key-13}. 

Experimental bound on $l_{8}$ mass presented in \cite{key-4}, namely,
$m_{l_{8}}$> 86 GeV is based on 25 years old CDF search for pair
production of unit-charged particles which leave the detector before
decaying \cite{key-14}. As mentioned in \cite{key-15} DO clearly
exclude 200 GeV leptogluons decaying within the detector and could
naively place the constraint $m_{l_{8}}$> 325 GeV. The twenty years
old H1 search for color octet electron has excluded the compositeness
scale $\Lambda$ < 3 TeV for $m_{e_{8}}$ \ensuremath{\approx} 100
GeV and $\Lambda$ < 240 GeV for $m_{e_{8}}$\ensuremath{\approx}
250 GeV \cite{key-16,key-17}. While the LEP experiments did not perform
dedicated search for leptogluons, low limits for excited lepton masses,
namely 103.2 GeV \cite{key-4}, certainly is valid for color octet
leptons, too. Finally, reconsideration of CMS results on leptoquark
searches performed in \cite{key-6} leads to the strongest limit $m_{e_{8}}$>
1.2-1.3 TeV. 

In this paper we analyze the potential of the FCC \cite{key-18} based
ep and $\mu$p colliders for charged leptogluon search. In Section
II we present main parameters of the FCC based lepton-hadron colliders.
Phenomenology of leptogluons is given in Section III. Resonant production
of color octet electrons at e-FCC and color octet muons at $\mu$-FCC
is analysed in Sections IV and V, respectively. In section VI achievable
values of compositeness scale are presented. Finally, section VII
contains summary of obtained results.

\section{FCC based $ep$ and $\mu p$ colliders}

FCC is future 100 TeV center-of-mass (CM) energy pp collider proposed
at CERN and supported by European Union within the Horizon 2020 Framework
Programme for Research and Innovation. It includes also an electron-positron
collider options at the same tunnel (TLEP), as well as ep collider
options. Construction of future $e^{+}e^{-}$ colliders (or special
e-linac) and $\mu^{+}\mu^{-}$ colliders tangential to FCC will give
opportunity to achieve highest CM energies in ep and $\mu$p collisions.
CM energy and luminosity values for different options are given in
Table 1. 

\begin{table}[H]
\caption{Main parameters of the FCC based lepton-hadron colliders \cite{key-19}}

\centering{}%
\begin{tabular}{|c|c|c|c|}
\hline 
Collider name  & $E_{l}$, TeV  & $\sqrt{s}$, TeV & $L_{int}=fb^{-1}$(per year) \tabularnewline
\hline 
\hline 
ERL60-FCC & 0.06 & 3.46 & 100\tabularnewline
\hline 
FCC-e80 & 0.08 & 4.00 & 230\tabularnewline
\hline 
FCC-e120 & 0.12 & 4.90 & 120\tabularnewline
\hline 
FCC-e175 & 0.175 & 5.92 & 40\tabularnewline
\hline 
OPL500-FCC & 0.5 & 10.0 & 10 - 100\tabularnewline
\hline 
OPERL500-FCC & 0.5 & 10.0 & 100 - 300\tabularnewline
\hline 
OPL1000-FCC & 1 & 14.1 & 5 - 50\tabularnewline
\hline 
OPERL1000-FCC & 1 & 14.1 & 50 - 150\tabularnewline
\hline 
OPL5000-FCC & 5 & 31.6 & 1 - 10\tabularnewline
\hline 
OPERL5000-FCC & 5 & 31.6 & 10 - 30\tabularnewline
\hline 
$\mu$63-FCC & 0.063 & 3.50 & 0.1 - 1\tabularnewline
\hline 
$\mu$175-FCC & 0.175  & 5.92 & 2 - 20\tabularnewline
\hline 
$\mu$750-FCC & 0.75  & 12.2 & 5 - 50\tabularnewline
\hline 
$\mu$1500-FCC & 1.5 & 17.3 & 5 - 50\tabularnewline
\hline 
$\mu$3000-FCC & 3 & 24.5 & 10 - 100\tabularnewline
\hline 
\end{tabular}
\end{table}

In Table 1 ERL60 denotes conventional energy recovery; e80, e120 and
e175 denote e-ring in the FCC tunnel; OPL denotes one pass linac tangential
to the FCC; OPERL denotes adding second (decelerating) linac shoulder
for energy recovery. Last 5 rows denote construction of $\mu$-rings
tangential to the FCC (for details see \cite{key-19}).

\section{Color Octet Leptons}

Following reference \cite{key-3} we assume that preons are color
triplets and follow usual statistics (Fermi-Dirac for fermions and
Bose-Einstein for bosons), which means that SM fermions should contain
odd number of fermionic preons. In fermion-scalar models leptons are
bound states of one fermionic preon and one scalar anti-preon 

\begin{equation}
l=(F\overline{S})=1+8
\end{equation}
therefore, each SM lepton has one color octet partner. In three-fermion
models the color decomposition is

\begin{equation}
l=(FFF)=1+8+8+10
\end{equation}
therefore, each SM lepton has two color octet and one color decuplet
partners. As for quark sector, each SM quark has anti-sextet partner
in fermion-scalar models (anti-triplet, anti-sextet and 15-plet partners
in three-fermion models). 

Concerning the relation between compositeness scale and masses of
leptogluons, two scenarios can be considered: $m_{l_{8}}$ \ensuremath{\approx}
$\Lambda$ (QCD-like scenario) and $m_{l_{8}}$ <\textcompwordmark{}<
$\Lambda$ (Higgs-like scenario). In the second scenario SM-like hierarchy
may be realized, namely, $m_{e_{8}}$ <\textcompwordmark{}< $m_{\mu_{8}}$<\textcompwordmark{}<
$m_{\tau8}$ <\textcompwordmark{}< $\Lambda$. Hereafter, numerical
calculations will be performed for $\Lambda$ = $m_{l_{8}}$ and $\Lambda$
=100 TeV cases. 

For the interaction of leptogluons with the corresponding lepton and
gluon we use the following Lagrangian \cite{key-4,key-9}:

\begin{equation}
L=\frac{1}{2\Lambda}\underset{l}{\sum}\left\{ \overline{l}_{g}^{\alpha}g_{s}G_{\mu\nu}^{\alpha}\sigma^{\mu\nu}(\eta_{L}l_{L}+\eta_{R}l_{R})+h.c.\right\} 
\end{equation}

where $G_{\mu\nu}^{\alpha}$ is the field strength tensor for gluon,
index $\alpha$ = 1, 2, ..., 8 denotes the color, $g_{s}$ is Gauge
coupling, $\eta_{L}$ and $\eta_{R}$ are the chirality factors, $l_{L}$
and $l_{R}$ denote left and right spinor components of lepton, $\sigma^{\mu\nu}$
is the antisymmetric tensor and $\Lambda$ is the compositeness scale.
The leptonic chiral invariance implies $\eta_{L}$$\eta_{R}$ = 0.
For numerical calculations we add leptogluons into the CalcHEP program
\cite{key-20}.

Decay width of the color octet lepton is given by

\begin{equation}
\Gamma(l_{8}\rightarrow l+g)=\frac{\alpha_{s}M_{l_{8}}^{3}}{4\Lambda^{2}}
\end{equation}

In Fig. 1 the decay width of leptogluons are presented for two scenarios,
namely, $\Lambda$ = $m_{l_{8}}$ and $\Lambda$ =100 TeV. The resonant
$l_{8}$ production cross sections for different options of the FCC
based lp colliders (Table I), evaluated using CalcHEP with CTEQ6L
parametrization \cite{key-21} for parton distribution functions,
are presented in Figs. 2 and 3 (for $\Lambda$ = $m_{l_{8}}$ and
$\Lambda$ =100 TeV, respectively). At this stage we ignore beamstrahlung
effects, which leads to reduction of cross sections at ep colliders
(see next section). 

\begin{figure}[H]
\begin{centering}
\includegraphics[scale=0.5]{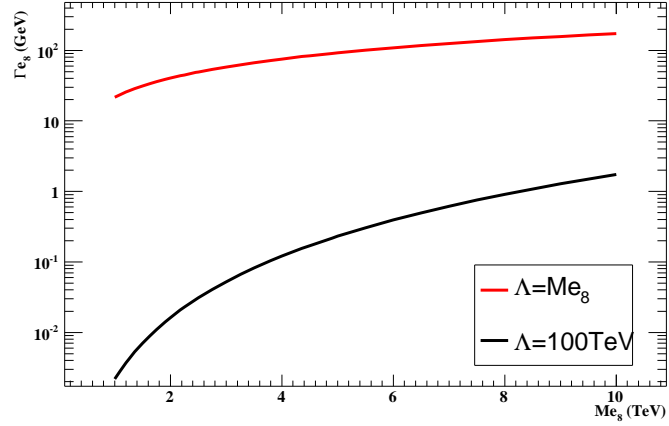}
\par\end{centering}

\centering{}\caption{Leptogluon decay width vs its mass for $\Lambda$ = $m_{l_{8}}$ and
$\Lambda$ =100 TeV. }
\end{figure}

\begin{figure}[H]
\begin{centering}
\includegraphics[scale=0.5]{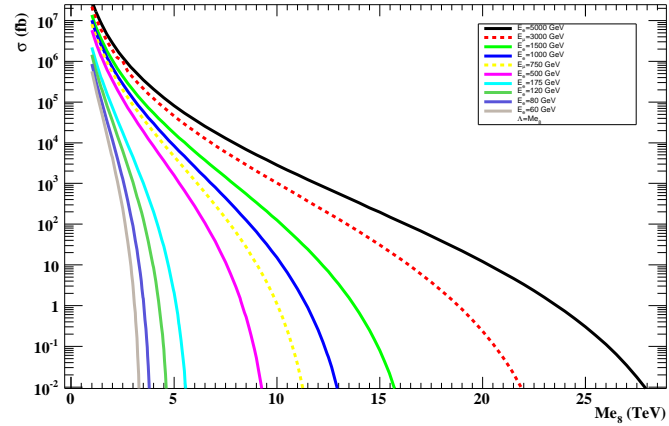}
\par\end{centering}

\centering{}\caption{Resonant $l_{8}$ production at the FCC based lp colliders for $\Lambda$
= $m_{l_{8}}$.}
\end{figure}

\begin{figure}[H]
\begin{centering}
\includegraphics[scale=0.5]{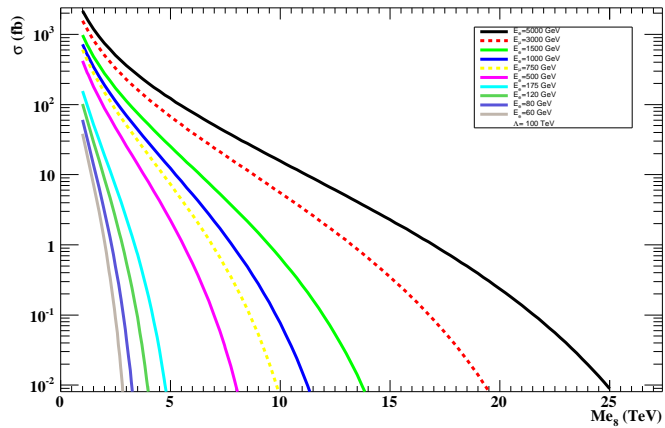}
\par\end{centering}

\centering{}\caption{Same as Fig. 2 for $\Lambda$ = 100 TeV.}
\end{figure}

\section{Color octet electrons at the FCC based ep colliders: Signal vs background}

In this case beamstrahlung reduces production cross-section of color
octet electrons, especially at large $m_{e_{8}}$ values. This reduction
is illustrated in Table 2 for ERL60-FCC. Analysis in this section
is performed taking into account beamstrahlung effects using \textquotedbl{}beamstrahlung
on\textquotedbl{} option for initial electron state in CalcHEP.

\begin{table}[H]
\caption{Effect of beamstrahlung at ERL60-FCC (cross-sections are given for
$\Lambda$ = $m_{e_{8}}$)}

\centering{}%
\begin{tabular}{|c|c|c|c|}
\hline 
\multirow{2}{*}{$m_{e_{8}}$, GeV} & \multicolumn{2}{c|}{Cross-section, fb} & \multirow{2}{*}{Reduction }\tabularnewline
\cline{2-3} 
 & Beamstrahlung on & Beamstrahlung off  & \tabularnewline
\hline 
\hline 
1000 & 3.59 x 10\textsuperscript{5} & 3.87 x 10\textsuperscript{5} & 0.93\tabularnewline
\hline 
2000 & 2.32 x 10\textsuperscript{3} & 2.67x 10\textsuperscript{3} & 0.87\tabularnewline
\hline 
3000 & 6.49 & 7.66 & 0.85\tabularnewline
\hline 
\end{tabular}
\end{table}

In order to determine appropriate cuts we start with consideration
of $p_{t}$ and $\eta$ distributions for signal and background processes.
Numerical calculations are performed at the partonic level using CalcHEP
simulation program \cite{key-20} with CTEQ6L parton distribution
functions \cite{key-21} and generic cuts $p_{t}(e)$ > 30 GeV, $p_{t}(j)$
> 50 GeV, where j means gluon for signal and quarks for background
processes. Main contributions to the background came from lepton-quark
scatterings via photon and Z-boson exchange. 

Transverse momentum distributions of final state jets for signal (with
$\Lambda$ = $m_{e_{8}}$) and background are shown in Fig. 4. Let
us mention that same distributions are valid for final electrons,
too. It is seen that $p_{t}$ > 400 GeV cut essentially reduces background,
whereas signal is almost unaffected (especially for large $m_{e_{8}}$
values). Below we use $p_{t}(e)$ > 400 GeV, $p_{t}(j)$ > 400 GeV
as a discovery cut for all ep colliders, keeping in mind $m_{e_{8}}$
> 1.2 TeV from the LHC $\sqrt{s}$ = 8 TeV data \cite{key-6}. Pseudo-rapidity
distributions for final electrons and jets are presented in Figs.
5 and 6, respectively. Corresponding discovery cuts for different
ep collider options are given in Table 3. 

\begin{figure}[H]
\begin{centering}
\includegraphics[scale=0.4]{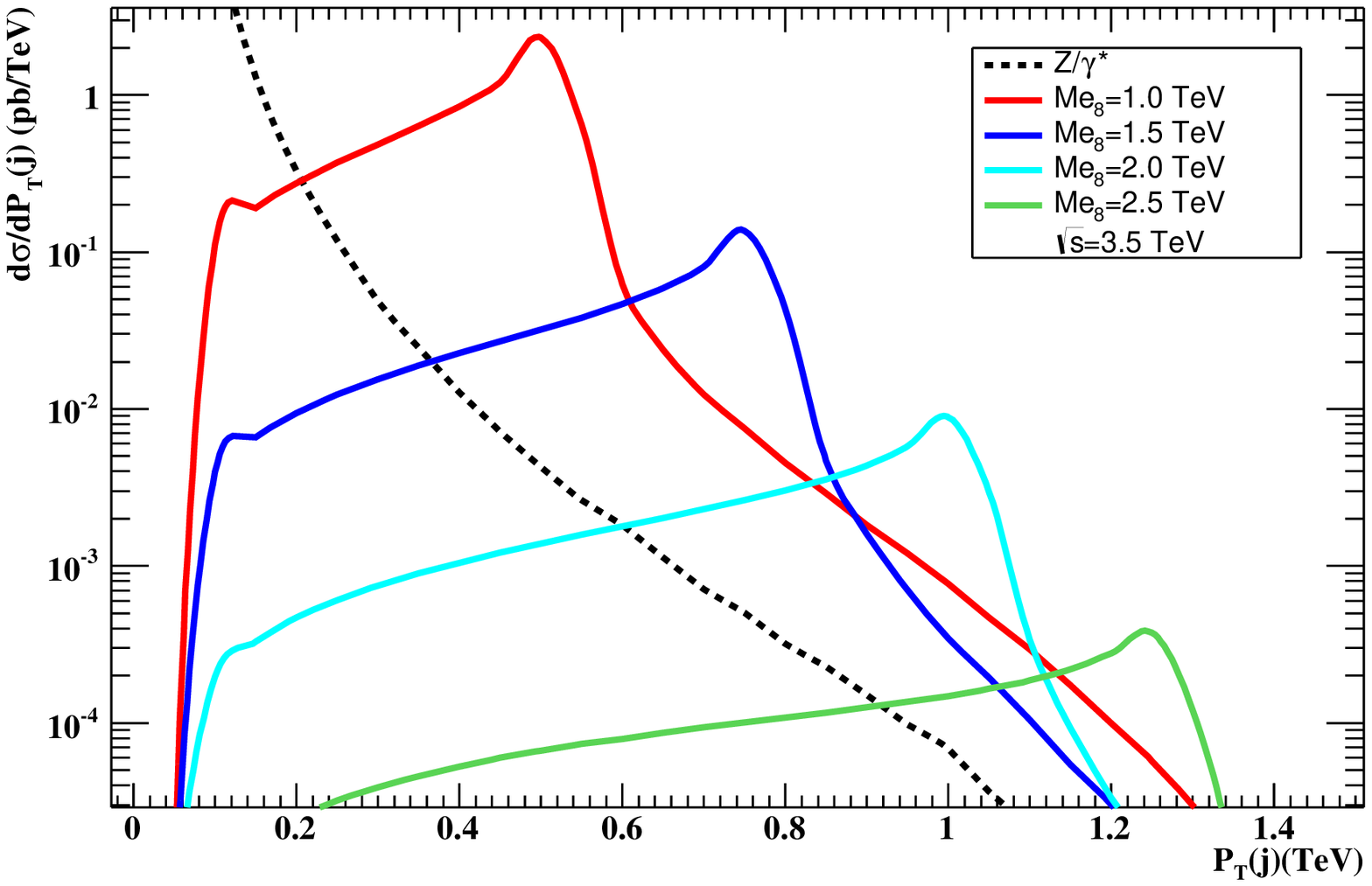}\includegraphics[scale=0.4]{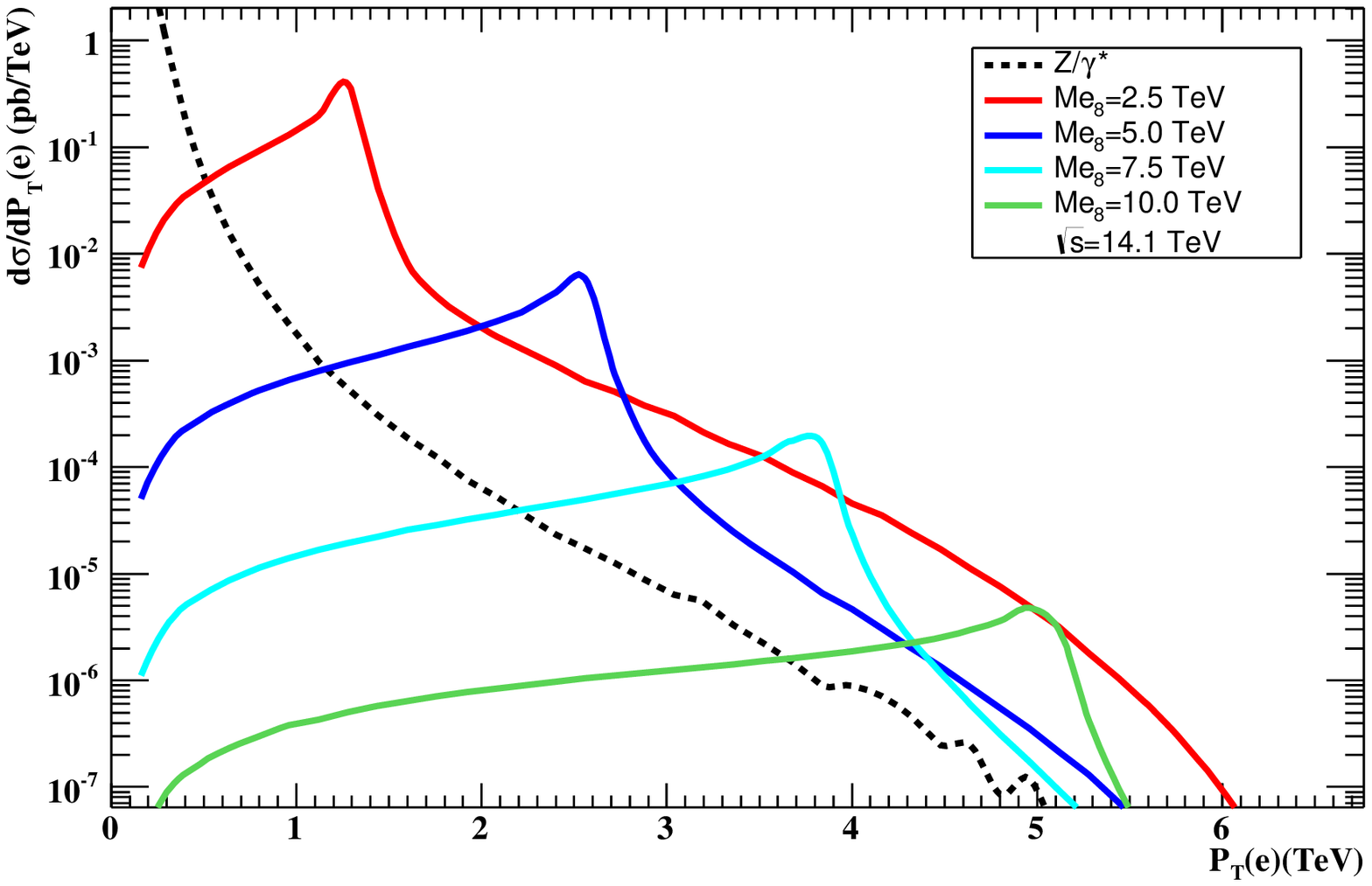}
\par\end{centering}

\centering{}\caption{Transverse momentum distributions of final state jets (and electrons)
for signal and background at ERL60-FCC (left) and OPL1000-FCC (right). }
\end{figure}

\begin{figure}[H]
\begin{centering}
\includegraphics[scale=0.4]{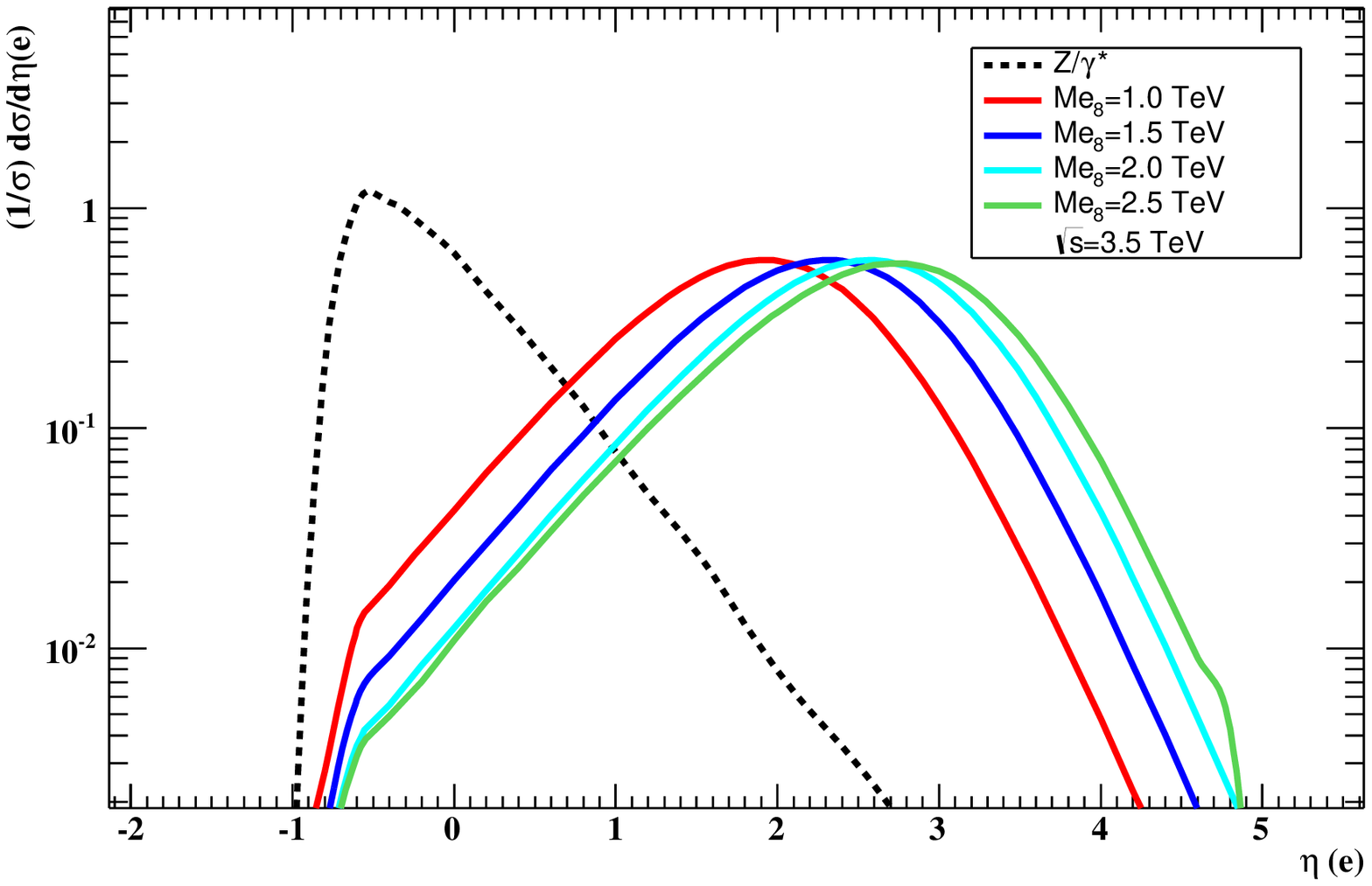}\includegraphics[scale=0.4]{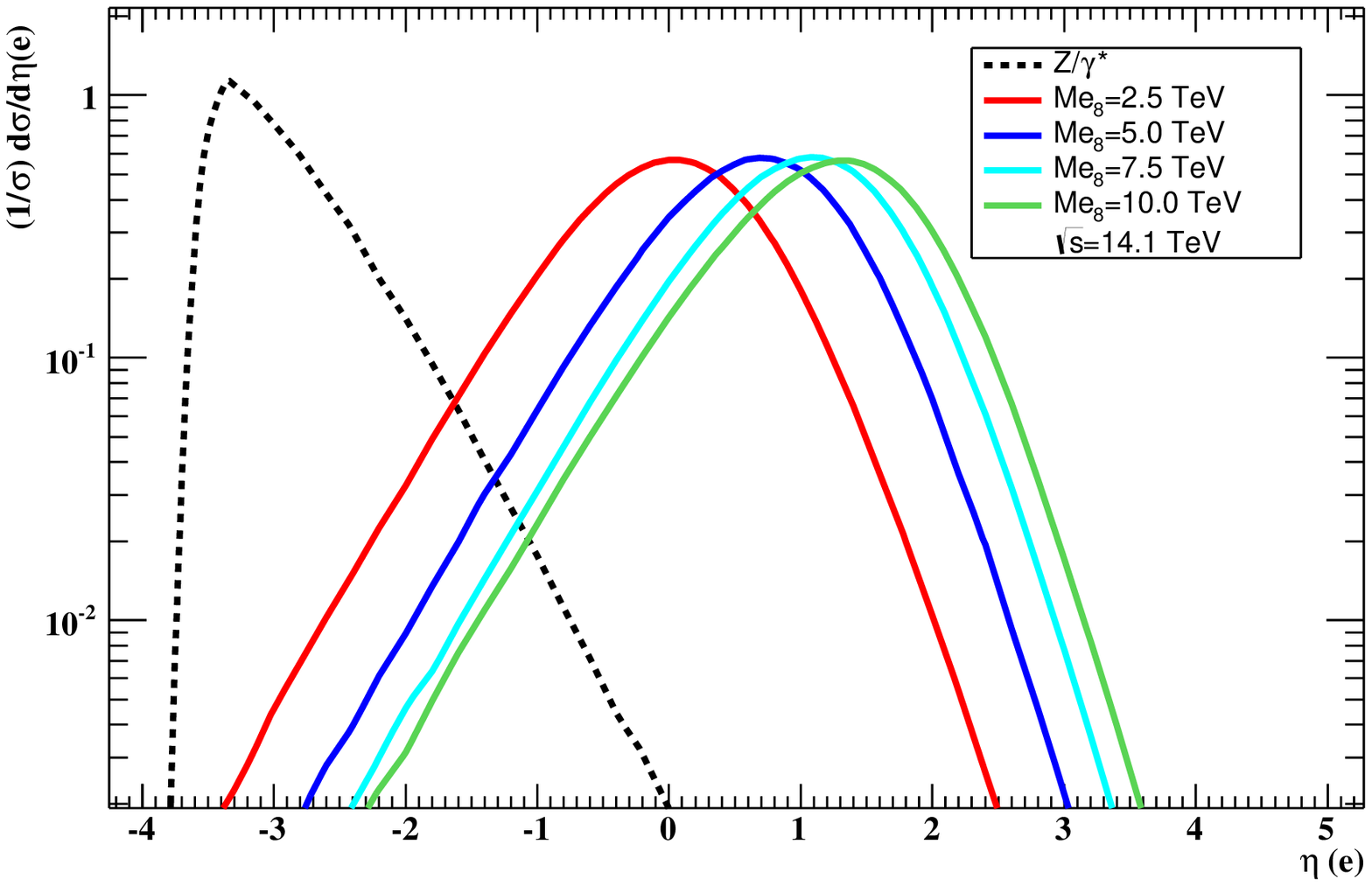}
\par\end{centering}

\centering{}\caption{Normalized pseudo-rapidity distributions of final electrons for signal
and background at ERL60-FCC (left) and OPL1000-FCC (right). }
\end{figure}

\begin{figure}[H]
\begin{centering}
\includegraphics[scale=0.4]{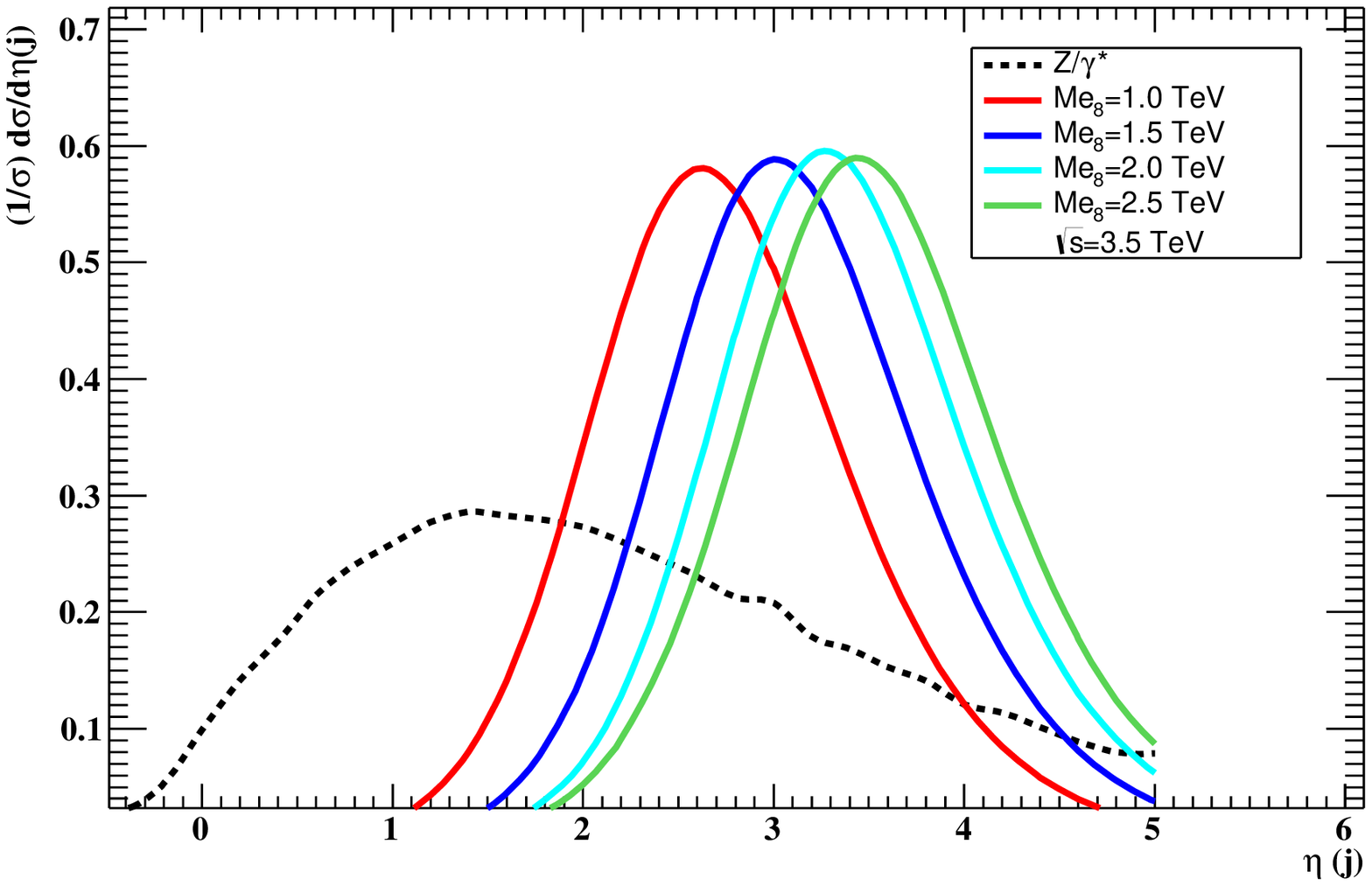}\includegraphics[scale=0.4]{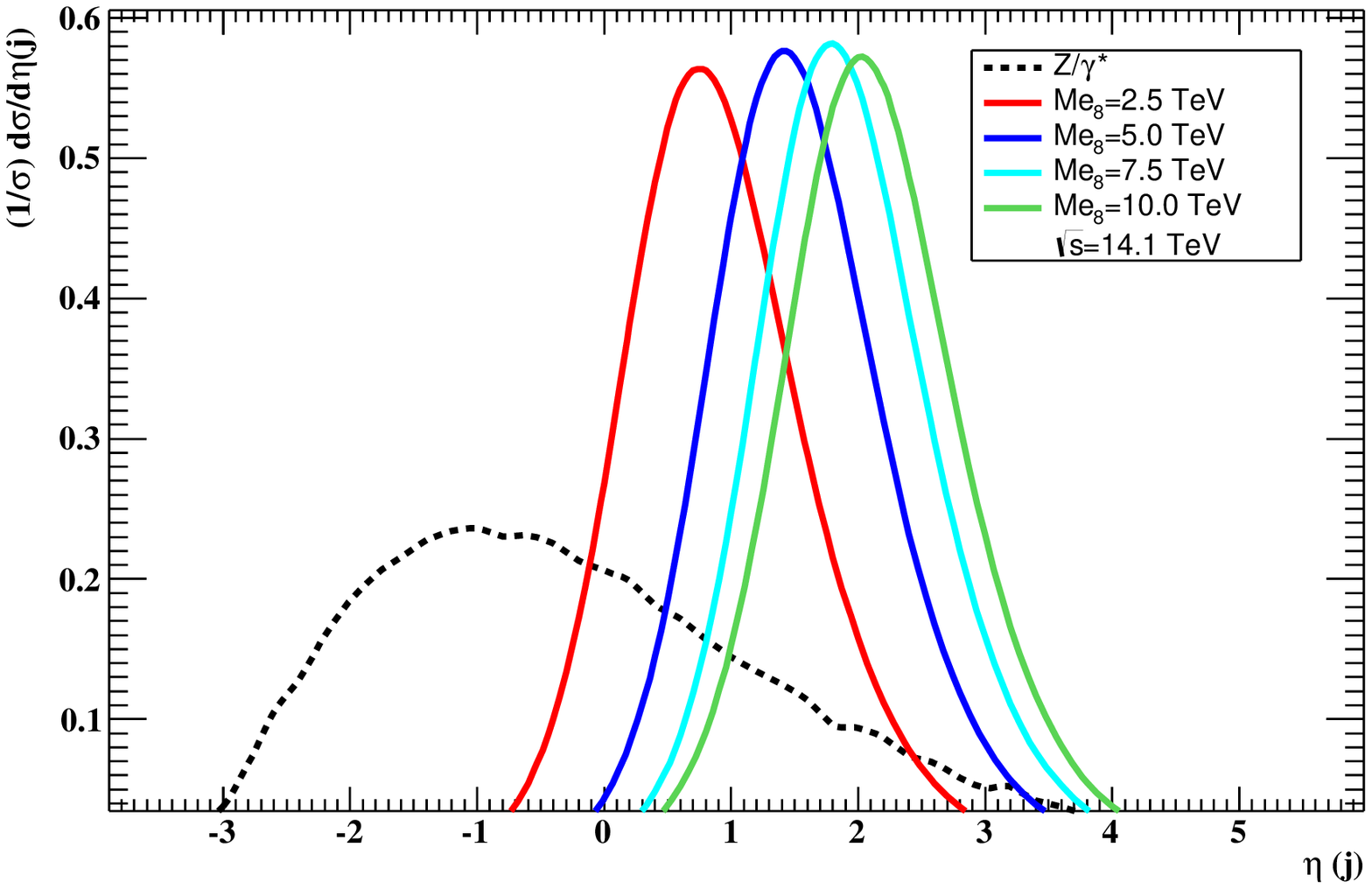}
\par\end{centering}

\centering{}\caption{Normalized pseudo-rapidity distributions of final jets for signal
and background at ERL60-FCC (left) and OPL1000-FCC (right). }
\end{figure}

\begin{table}[H]
\caption{Pseudorapidity cuts for different ep collider options }

\centering{}%
\begin{tabular}{|c|c|c|c|c|}
\hline 
\multirow{1}{*}{Electron Energy, GeV} & \multirow{1}{*}{60} & 175 & 500 & 1000\tabularnewline
\hline 
\hline 
$\eta_{e}$ & 1<$\eta_{e}$<4 & 0<$\eta_{e}$<4 & -1<$\eta_{e}$<4 & -1.5<$\eta_{e}$<4\tabularnewline
\hline 
$\eta_{j}$ & 2<$\eta_{j}$<4 & 1<$\eta_{j}$<4 & 0<$\eta_{j}$<4 & -0.5<$\eta_{j}$<4\tabularnewline
\hline 
\end{tabular}
\end{table}

In Table 4 we present observation (3$\sigma$) and discovery (5$\sigma$)
limits on masses of color octet electrons for different FCC based
ep collider options. For statistical significance we use 

\begin{equation}
S=\frac{\sigma_{s}}{\sqrt{\sigma_{s}+\sigma_{b}}}\sqrt{L_{int}}
\end{equation}
where $\sigma_{s}$ ($\sigma_{_{b}}$) means signal (background) cross
section and $L_{int}$ is integrated luminosity. With the pair production
channel, the 14 TeV LHC can probe color octet electrons with masses
up to 2.5 TeV with 100 fb\textsuperscript{-1}of integrated luminosity
\cite{key-5}. It is seen that FCC based ep colliders cover essentially
wider region of $e_{8}$'s mass. 

\begin{table}[H]
\caption{Observation (3$\sigma$) and discovery (5$\sigma$) limits for color
octet electrons }

\centering{}%
\begin{tabular}{|c|c|c|c|c|}
\hline 
\multirow{2}{*}{Collider Name} & \multirow{2}{*}{$\Lambda$} & \multirow{2}{*}{$L_{int}$, fb\textsuperscript{-1}} & \multicolumn{2}{c|}{$m_{e_{8}}$, GeV}\tabularnewline
\cline{4-5} 
 &  &  & 3$\sigma$ & 5$\sigma$\tabularnewline
\hline 
\multirow{4}{*}{ERL60-FCC $\sqrt{s}$ = 3.46 TeV} & \multirow{2}{*}{$m_{e_{8}}$} & 10 & 2990 & 2900\tabularnewline
\cline{3-5} 
 &  & 100 & 3150 & 3085\tabularnewline
\cline{2-5} 
 & \multirow{2}{*}{100 TeV} & 10 & 1150 & -\tabularnewline
\cline{3-5} 
 &  & 100 & 1690 & 1485\tabularnewline
\hline 
\multirow{2}{*}{FCC-e175 $\sqrt{s}$ = 5.92 TeV} & $m_{e_{8}}$ & 40 & 5110 & 4970\tabularnewline
\cline{2-5} 
 & 100 TeV & 40 & 2675 & 2350\tabularnewline
\hline 
\multirow{4}{*}{OPL500-FCC $\sqrt{s}$ = 10.0 TeV} & \multirow{2}{*}{$m_{e_{8}}$} & 10 & 7825 & 7500\tabularnewline
\cline{3-5} 
 &  & 100 & 8450 & 6600\tabularnewline
\cline{2-5} 
 & \multirow{2}{*}{100 TeV} & 10 & 3800 & 3200\tabularnewline
\cline{3-5} 
 &  & 100 & 5070 & 4520\tabularnewline
\hline 
\multirow{4}{*}{OPL1000-FCC $\sqrt{s}$ = 14.1 TeV} & \multirow{2}{*}{$m_{e_{8}}$} & 5 & 10200 & 9640\tabularnewline
\cline{3-5} 
 &  & 50 & 11220 & 10800\tabularnewline
\cline{2-5} 
 & \multirow{2}{*}{100 TeV} & 5 & 5000 & 4100\tabularnewline
\cline{3-5} 
 &  & 50 & 6750 & 6000\tabularnewline
\hline 
\end{tabular}
\end{table}

\section{Color octet muons at the FCC based \textmu{}p colliders: Signal vs
background}

For illustration we consider $\mu$750-FCC. Transverse momentum distributions
of final state muons for signal (with $\Lambda=m_{e_{8}}$) and background
are shown in Fig. 7. Let us remind that same distributions are valid
for final state jets, too. Similar to ep case, we use $p_{t}(\mu)$
> 400 GeV, $p_{t}(j)$ > 400 GeV as a discovery cut for all $\mu$p
colliders (rough estimations show that color octet muons with masses
below 1 TeV are excluded by the LHC $\sqrt{s}$ = 8 TeV data). Pseudo-rapidity
distributions for final muons and jets are presented in Figs. 8 and
9, respectively. Corresponding discovery cuts for different $\mu$p
collider options are given in Table 5.

\begin{figure}[H]
\begin{centering}
\includegraphics[scale=0.5]{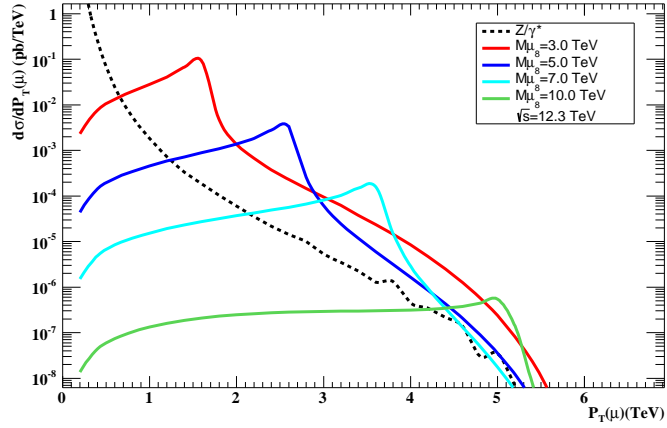}
\par\end{centering}

\centering{}\caption{Transverse momentum distributions of final state muons (and jets)
for signal and background at $\mu$750-FCC. }
\end{figure}

\begin{figure}[H]
\begin{centering}
\includegraphics[scale=0.5]{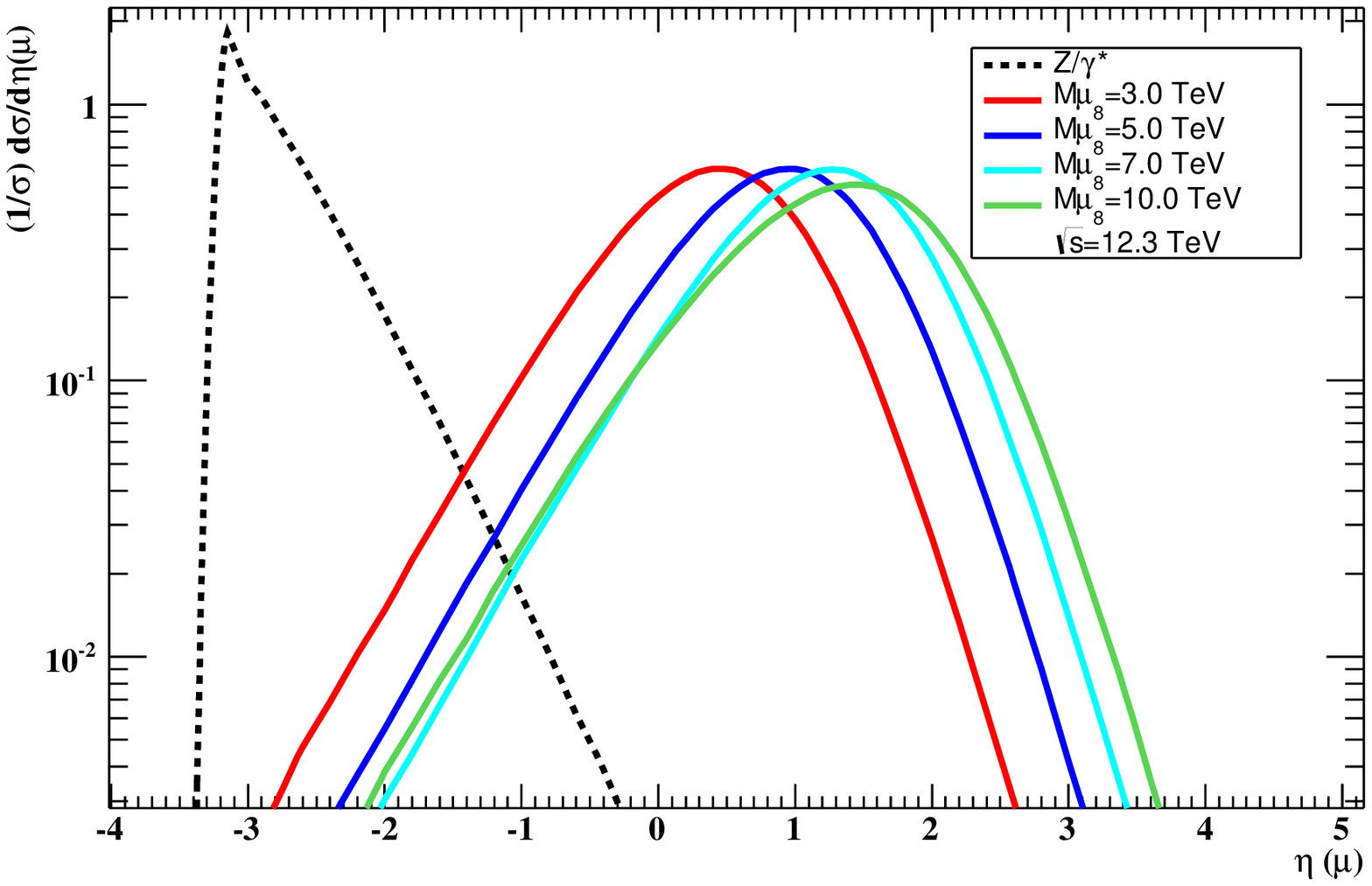}
\par\end{centering}

\centering{}\caption{Normalized pseudo-rapidity distributions of final muons for signal
and background at $\mu$750-FCC. }
\end{figure}

\begin{figure}[H]
\begin{centering}
\includegraphics[scale=0.5]{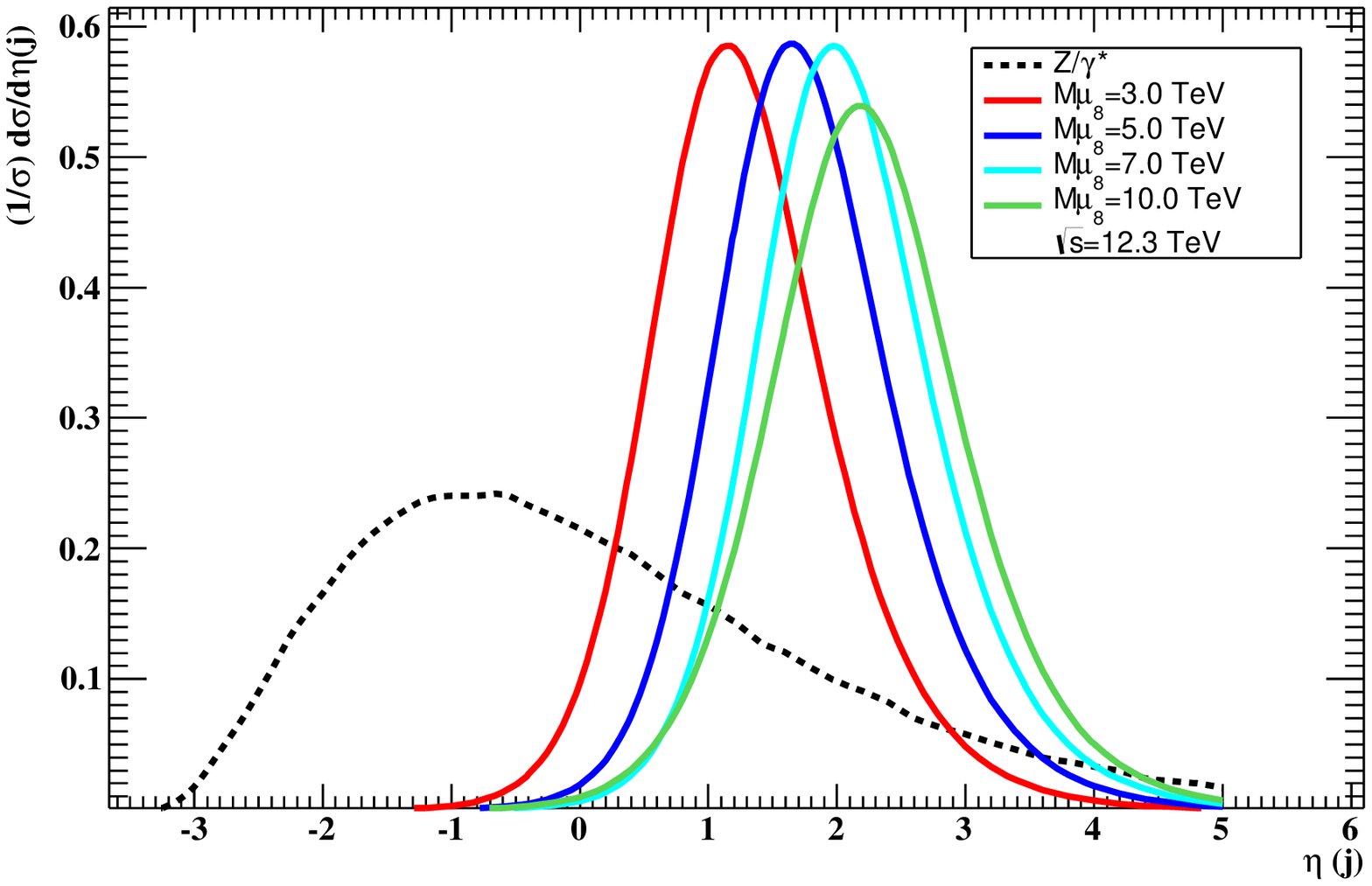}
\par\end{centering}

\centering{}\caption{Normalized pseudo-rapidity distributions of final jets for signal
and background at $\mu$750-FCC. }
\end{figure}

\begin{table}[H]
\caption{Pseudorapidity cuts for different $\mu$p collider options }

\centering{}%
\begin{tabular}{|c|c|c|c|}
\hline 
\multirow{1}{*}{Muon Energy, GeV} & \multirow{1}{*}{63} & 175 & 750\tabularnewline
\hline 
\hline 
$\eta_{\mu}$ & 1<$\eta_{\mu}$<4 & 0<$\eta_{\mu}$<4 & -1<$\eta_{\mu}$<4\tabularnewline
\hline 
$\eta_{j}$ & 2<$\eta_{j}$<4 & 1<$\eta_{j}$<4 & 0<$\eta_{j}$<4\tabularnewline
\hline 
\end{tabular}
\end{table}

In Table 6 we present observation (3$\sigma$) and discovery (5$\sigma$)
limits on masses of color octet muons for different FCC based $\mu$p
collider options. Again FCC based $\mu$p colliders cover essentially
wider region of $\mu_{8}$'s mass comparing to the LHC.

\begin{table}[H]
\caption{Observation (3$\sigma$) and discovery (5$\sigma$) limits for color
octet muons }

\centering{}%
\begin{tabular}{|c|c|c|c|c|}
\hline 
\multirow{2}{*}{Collider Name} & \multirow{2}{*}{$\Lambda$} & \multirow{2}{*}{$L_{int}$, fb\textsuperscript{-1}} & \multicolumn{2}{c|}{$m_{\mu_{8}}$, GeV}\tabularnewline
\cline{4-5} 
 &  &  & 3$\sigma$ & 5$\sigma$\tabularnewline
\hline 
\multirow{4}{*}{\textmu{}63-FCC $\sqrt{s}$= 3.50 TeV} & \multirow{2}{*}{$m_{\mu_{8}}$} & 0.1 & 2580 & 2430\tabularnewline
\cline{3-5} 
 &  & 1 & 2880 & 2760\tabularnewline
\cline{2-5} 
 & \multirow{2}{*}{100 TeV} & 0.1 & - & -\tabularnewline
\cline{3-5} 
 &  & 1 & - & -\tabularnewline
\hline 
\multirow{4}{*}{\textmu{}175-FCC $\sqrt{s}$ = 5.92 TeV} & \multirow{2}{*}{$m_{\mu_{8}}$} & 2 & 4700 & 4500\tabularnewline
\cline{3-5} 
 &  & 20 & 5080 & 4930\tabularnewline
\cline{2-5} 
 & \multirow{2}{*}{100 TeV} & 2 & 1450 & -\tabularnewline
\cline{3-5} 
 &  & 20 & 2230 & 1900\tabularnewline
\hline 
\multirow{4}{*}{\textmu{}750-FCC $\sqrt{s}$= 12.2 TeV} & \multirow{2}{*}{$m_{\mu_{8}}$} & 5 & 9225 & 8780\tabularnewline
\cline{3-5} 
 &  & 50 & 10060 & 9730\tabularnewline
\cline{2-5} 
 & \multirow{2}{*}{100 TeV} & 5 & 3900 & 3200\tabularnewline
\cline{3-5} 
 &  & 50 & 5350 & 4700\tabularnewline
\hline 
\end{tabular}
\end{table}

\section{Compositeness scale}

Although the FCC based lp colliders will cover much wider $m_{l_{8}}$
mass regions than LHC, this is not the case for the FCC pp option:
rough estimations show that FCC-pp will give opportunity to discover
color octet leptons up to 20 TeV mass values. In this section we analyze
potential of e-FCC and $\mu$-FCC for determination of compositeness
scale. While the knowledge of color octet electron (and/or muon) mass
will give opportunity to further optimization of cuts for purpose
of $\Lambda$ determination, in our analysis we use $p_{t},\eta_{e},\eta_{j},\eta_{\mu},m_{inv}(lj)$
cut values given in sections VI and V for $e_{8}$ and $\mu_{8}$
, respectively. Achievable compositeness scales at the FCC based ep
and $\mu$p colliders are presented in Tables 7 and 8, respectively. 

\begin{table}[H]
\caption{Achievable compositeness scale in TeV units at the FCC based ep colliders. }

\centering{}%
\begin{tabular}{|c|c|c|c|c|}
\hline 
\multirow{2}{*}{ERL60-FCC} & \multicolumn{2}{c|}{3$\sigma$} & \multicolumn{2}{c|}{5$\sigma$}\tabularnewline
\cline{2-5} 
 & L=10 fb\textsuperscript{-1} & L=100 fb\textsuperscript{-1} & L=10 fb\textsuperscript{-1} & L=100 fb\textsuperscript{-1}\tabularnewline
\hline 
1000 & 100000 & 195000 & 85000 & 150000\tabularnewline
\hline 
1500 & 62000 & 105000 & 49000 & 82000\tabularnewline
\hline 
2000 & 32000 & 51000 & 26800 & 48000\tabularnewline
\hline 
2500 & 15000 & 27000 & 10000 & 20000\tabularnewline
\hline 
\multirow{2}{*}{FCC-e175} & \multicolumn{2}{c|}{3$\sigma$} & \multicolumn{2}{c|}{5$\sigma$}\tabularnewline
\cline{2-5} 
 & \multicolumn{4}{c|}{L=40 fb\textsuperscript{-1}}\tabularnewline
\hline 
1000 & \multicolumn{2}{c|}{280000} & \multicolumn{2}{c|}{210000}\tabularnewline
\hline 
2000 & \multicolumn{2}{c|}{135000} & \multicolumn{2}{c|}{122200}\tabularnewline
\hline 
3000 & \multicolumn{2}{c|}{60000} & \multicolumn{2}{c|}{47200}\tabularnewline
\hline 
4000 & \multicolumn{2}{c|}{27500} & \multicolumn{2}{c|}{21000}\tabularnewline
\hline 
\multirow{2}{*}{OPL500-FCC} & \multicolumn{2}{c|}{3$\sigma$} & \multicolumn{2}{c|}{5$\sigma$}\tabularnewline
\cline{2-5} 
 & L=10 fb\textsuperscript{-1} & L=100 fb\textsuperscript{-1} & L=10 fb\textsuperscript{-1} & L=100 fb\textsuperscript{-1}\tabularnewline
\hline 
1000 & 363000 & 653000 & 277000 & 503000\tabularnewline
\hline 
3000 & 156250 & 283000 & 119000 & 218000\tabularnewline
\hline 
5000 & 57500 & 105500 & 43250 & 81000\tabularnewline
\hline 
7000 & 16750 & 32000 & 12000 & 24000\tabularnewline
\hline 
\multirow{2}{*}{OPL1000-FCC} & \multicolumn{2}{c|}{3$\sigma$} & \multicolumn{2}{c|}{5$\sigma$}\tabularnewline
\cline{2-5} 
 & L=5 fb\textsuperscript{-1} & L=50 fb\textsuperscript{-1} & L=5 fb\textsuperscript{-1} & L=50 fb\textsuperscript{-1}\tabularnewline
\hline 
1000 & 255000 & 368000 & 191000 & 342000\tabularnewline
\hline 
2500 & 172500 & 295000 & 126000 & 228000\tabularnewline
\hline 
5000 & 67000 & 120000 & 52000 & 97000\tabularnewline
\hline 
7500 & 29000 & 54000 & 22000 & 41000\tabularnewline
\hline 
10000 & 11420 & 23000 & 7750 & 16750\tabularnewline
\hline 
\end{tabular}
\end{table}

\begin{table}[H]
\caption{Achievable compositeness scale in TeV units at the FCC based $\mu$p
colliders. }

\centering{}%
\begin{tabular}{|c|c|c|c|c|}
\hline 
\multirow{2}{*}{FCC-$\mu$175} & \multicolumn{2}{c|}{3$\sigma$} & \multicolumn{2}{c|}{5$\sigma$}\tabularnewline
\cline{2-5} 
 & L=2 fb\textsuperscript{-1} & L=20 fb\textsuperscript{-1} & L=2 fb\textsuperscript{-1} & L=20 fb\textsuperscript{-1}\tabularnewline
\hline 
1000 & 129000 & 234000 & 98000 & 180000\tabularnewline
\hline 
2000 & 66250 & 119250 & 50000 & 92000\tabularnewline
\hline 
3000 & 29750 & 54500 & 22250 & 41750\tabularnewline
\hline 
4000 & 13250 & 25750 & 9500 & 19250\tabularnewline
\hline 
\multirow{2}{*}{$\mu$750-FCC} & \multicolumn{2}{c|}{3$\sigma$} & \multicolumn{2}{c|}{5$\sigma$}\tabularnewline
\cline{2-5} 
 & L=5 fb\textsuperscript{-1} & L=50 fb\textsuperscript{-1} & L=5 fb\textsuperscript{-1} & L=50 fb\textsuperscript{-1}\tabularnewline
\hline 
1000 & \multicolumn{1}{c|}{264000} & 474000 & \multicolumn{1}{c|}{203000} & 367000\tabularnewline
\hline 
3000 & 141000 & 254000 & 108000 & 196000\tabularnewline
\hline 
5000 & 63500 & 114500 & 48250 & 88250\tabularnewline
\hline 
7000 & 27500 & 50750 & 20750 & 39000\tabularnewline
\hline 
10000 & 4500 & 10750 & 1250 & 8000\tabularnewline
\hline 
\end{tabular}
\end{table}

\section{Conclusions}

Certainly, if color octet leptons have mass values covered by the
FCC based lp colliders, they will be observed earlier at the FCC pp
option. Nevertheless, e-FCC and $\mu$-FCC will give opportunity to
obtain very important information which cannot be handled by the FCC-pp.
As shown in section VI, compositeness scale well above 100 TeV can
be probed. Very important feature of OPL-FCC ep colliders, namely,
longitudinal polarization of electrons will give opportunity to determine
the Lorentz structure of $l_{8}$-l-g vertex (the work on the subject
is under progress). In general, lepton-hadron colliders has cleaner
environment than hadron colliders. Finally, possible discovery of
color octet leptons at the FCC-pp will determine the type of future
lp collider to be installed.
\begin{acknowledgments}
This study is supported by TUBITAK under the grant no 114F337.\end{acknowledgments}


\begin{thebibliography}{References}
\bibitem{key-1}I.A. D'Souza and C.S. Kalman, \textquotedblleft{}PREONS:
Models of Leptons, Quarks and Gauge Bosons as Composite Objects\textquotedblright{},
World Scientific (1992).

\bibitem{key-2}H. Fritzsch and G. Mandelbaum, \textquotedblleft{}Weak
interactions as manifestations of the substructure of leptons and
quarks\textquotedblright{}, Phys. Lett. B 102 (1981) 319.

\bibitem{key-3}A. Celikel, M. Kantar and S. Sultansoy, \textquotedblleft{}A
search for sextet quarks and leptogluons at the LHC\textquotedblright{},
Phys. Lett. B 443 (1998) 359.

\bibitem{key-4}K.A. Olive et al. (Particle Data Group), Chin. Phys.
C 38 (2014) 090001.

\bibitem{key-5}T. Mandal and S. Mitra, \textquotedblleft{}Probing
color octet electrons at the LHC\textquotedblright{}, Phys. Rev. D
87 (2013) 095008.

\bibitem{key-6}D. Gonçalves-Netto et al., \textquotedblleft{}Looking
for leptogluons\textquotedblright{}, Phys. Rev. D 87 (2013) 094023.

\bibitem{key-7}T. Jelinski and D. Zhuridov, \textquotedblleft{}Leptogluons
in dilepton production at LHC\textquotedblright{}, e-Print: arXiv:1510.04872
{[}hep-ph{]}.

\bibitem{key-8}A. Celikel and M. Kantar, \textquotedblleft{}Resonance
Production of New Resonances at ep and $\gamma$p Colliders\textquotedblright{},
Tr. J. of Physics 22 (1998) 401.

\bibitem{key-9}M. Sahin, S. Sultansoy and S. Turkoz, \textquotedblleft{}Resonant
production of color octet electron at the LHeC\textquotedblright{},
Phys. Lett. B 689 (2010) 172.

\bibitem{key-10}M. Sahin, \textquotedblleft{}Resonant production
of spin-3/2 color octet electron at the LHeC\textquotedblright{},
Acta Physica Polonica B 45 (2014) 1811.

\bibitem{key-11}A.N. Akay, H. Karadeniz, M. Sahin and S. Sultansoy,
\textquotedblleft{}Indirect search for color octet electron at next-generation
linear colliders\textquotedblright{}, EPL 95 (2011) 31001.

\bibitem{key-12}K. Cheung, \textquotedblleft{}Muon-proton colliders:
Leptoquarks, contact interactions and extra dimensions\textquotedblright{},
AIP Conference Proceedings 542 (2000) 160.

\bibitem{key-13}A.N. Akay et al., \textquotedblleft{}New IceCube
data and color octet neutrino interpretation of the PeV energy events\textquotedblright{},
Int. J. Mod. Phys. A 30 (2015) 1550163.

\bibitem{key-14}F. Abe et al. (CDF Collaboration), \textquotedblleft{}Search
for Heavy Stable Charged Particles in 1.8-TeV pp(bar) collisions at
the Fermilab Collider\textquotedblright{}, Phys. Rev. Lett. 63 (1989)
1447.

\bibitem{key-15}J.L. Hewett and T.G. Rizzo, \textquotedblleft{}Much
ado about leptoquarks: A comprehensive analysis\textquotedblright{},
Phy. Rev. D 56 (1997) 5709.

\bibitem{key-16}I. Abt et al. (H1 Collaboration), \textquotedblleft{}A
search for leptoquarks, leptogluons and excited leptons in H1 at HERA\textquotedblright{},
Nucl. Phys. B 396 (1993) 3.

\bibitem{key-17}T. Ahmed et al. (H1 Collaboration), \textquotedblleft{}A
search for leptoquarks and squarks at HERA\textquotedblright{}, Z.
Phys. C 64 (1994) 545. 

\bibitem{key-18}FCC web page: https://fcc.web.cern.ch

\bibitem{key-19}Y.C. Acar, U. Kaya, B.B. Oner and S. Sultansoy, \textquotedblleft{}FCC
based ep and $\mu$p colliders\textquotedblright{}, arXiv:1510.08284
{[}hep-ex{]}

\bibitem{key-20}A. Belyayev, N.D. Christensen and A.Pukhov, \textquotedblleft{}CalcHEP
3.4 for collider physics within and beyond the Standard Model\textquotedblright{},
Comput. Phys. Commun. 184 (2013) 1729.

\bibitem{key-21}D. Stump et al., \textquotedblleft{}Inclusive jet
production, parton distributions and the search for new physics\textquotedblright{},
JHEP 0310 (2003) 046.\end{thebibliography}
\end{document}